\documentclass{PoS}
\pdfoutput=1
\usepackage{subfigure}
\usepackage{latexsym}
\usepackage{amsfonts}
\usepackage[latin1]{inputenc}
\usepackage{graphicx}
\usepackage{mathrsfs}
\usepackage{amssymb}
\usepackage{amsmath}
\usepackage{verbatim}
\usepackage{multirow}
\usepackage{soul}

\newcommand{\GeV}{{\ensuremath\rm GeV}}
\newcommand{\TeV}{{\ensuremath\rm TeV}}
\newcommand{\al}{\alpha}
\newcommand{\be}{\beta}
\newcommand{\lam}{\lambda}

\newcommand{\HS}{\texttt{HiggsSignals}}
\newcommand{\HB}{\texttt{HiggsBounds}}

\newcommand{\pstar}{\ensuremath{p_*}}

\newcommand{\lb}{\left (}
\newcommand{\rb}{\right )}
\def\noi{\noindent}

\title{The Higgs singlet extension at LHC Run 2}

\ShortTitle{Singlet LHC Run 2}
\author{{G. Chalons}\\
        Laboratoire de Physique Subatomique et de Cosmologie\\
Universit\'e Grenoble-Alpes, CNRS/IN2P3\\
53 Rue des Martyrs\\
F-38026 Grenoble Cedex, France\\
        E-mail: \email{chalons@lpsc.in2p3.fr}}
\author{D. L\'opez-Val\\
Center for Cosmology, Particle Physics and Phenomenology CP3\\
Universit\'e catholique de Louvain\\
Chemin du Cyclotron 2, B-1348 Louvain-la-Neuve, Belgium\\
        E-mail: \email{david.lopezval@uclouvain.be}}
\author{\speaker{T. Robens}\\
IKTP, Technische Universit\"at Dresden\\
Zellescher Weg 19, D-01069 Dresden, Germany\\
        E-mail: \email{Tania.Robens@tu-dresden.de}}
\author{T. Stefaniak\\
Department of Physics and Santa Cruz Institute for Particle Physics\\
University of California, Santa Cruz, CA 95064, USA\\
E-mail: \email{tistefan@ucsc.edu}}

\abstract{We discuss the current status of theoretical and experimental constraints on the real Higgs singlet extension of the Standard Model. For the second neutral (non-standard) Higgs boson the mass range up to 
1 \TeV{\,} accessible at past and current collider experiments is considered. We furthermore discuss electroweak corrections to the $H \,\rightarrow\, hh$ partial decay width within this model.\\ 
\vspace{2mm}\\
\rightline{CP3-16-34, LPSC16110}
}
\FullConference{XXIV International Workshop on Deep-Inelastic Scattering and Related Subjects\\
		11-15 April, 2016\\
		DESY Hamburg, Germany}

\begin{document}

\section{The model}
In this work we consider the simplest extension of the Standard Model (SM) Higgs sector, where an additional real {scalar} field is added \cite{Schabinger:2005ei, Patt:2006fw, Bowen:2007ia}. The model contains a complex $SU(2)_L$ doublet, in the following denoted by $\Phi$, and a real scalar $S$ which is a singlet under the SM gauge group. The most general renormalizable Lagrangian compatible with an additional $Z_2$ symmetry is then given by
$
\mathscr{L}_s = \left( D^{\mu} \Phi \right) ^{\dagger} D_{\mu} \Phi + 
\partial^{\mu} S \partial_{\mu} S - V(\Phi,S ),$
with the scalar potential
\begin{eqnarray}\label{potential}
V(\Phi,S ) \,=\, -m^2 \Phi^{\dagger} \Phi -\mu ^2 S ^2 + \lambda_1
(\Phi^{\dagger} \Phi)^2 + \lambda_2  S^4 + \lambda_3 \Phi^{\dagger}
\Phi S ^2.
\end{eqnarray}
In the unitary gauge, the Higgs fields are given by
$\Phi \equiv
\left(
0 \;\;
\tfrac{\tilde{h}+v}{\sqrt{2}}
\right)^T, \,
S \equiv \frac{h'+v_s}{\sqrt{2}}$, with $v,\,v_s$ denoting the non-zero vacuum expectation values of the doublet and singlet.
 Physically, the above potential leads to a mixing between the gauge eigenstates, related via the mixing angle $\al$ according to $h\,=\,c_\al\,\tilde{h}-s_\al\,h',\,H\,=\,s_\al\,\tilde{h}+c_\al\,h'$, where we used the shorthand notation $s_\al\,(c_\al)\,\equiv\,\sin\al \lb\cos\al \rb$.
We here use the convention that $m_h\,\leq\,m_H$, and choose {as input
parameters}
$m_h,\,m_H,\,\sin\alpha,\,v,\,\tan\beta\,\equiv\,\frac{v}{v_s}$,
where $v\,\sim\,246\,\GeV$. In addition, one of the scalar masses is fixed to $\sim\,125\,\GeV${, where we distinguish between} the {\sl high-mass} ($m_h\,\sim\,m_{h,\text{SM}}$) and {\sl low-mass} ($m_H\,\sim\,m_{h,\text{SM}}$) scenario.
The above mixing also leads to {the familiar} rescaling of the {SM-like Higgs} couplings at tree level by $\sin\al \,\lb \cos\al \rb$ for {h(H)}, with respect to the couplings for a SM Higgs boson of that mass.

\section{Parameter constraints and predictions at the LHC Run 2}

We refer the reader to \cite{Pruna:2013bma,Robens:2015gla,Robens:2016xkb} for a detailed discussion of the individual constraints. Vacuum stability, perturbative unitarity, perturbativity of the couplings, agreement with electroweak precision observables have been explicitly discussed in the above references; constraints from the $W$-boson mass measurement follow \cite{Lopez-Val:2014jva}. In \cite{Robens:2016xkb}, previous results were updated {especially with regard to the latest LHC limits and Higgs signal strength measurements \cite{ATLAS-CONF-2015-044}, using the public tools \HB\ (version 4.3.1)~\cite{Bechtle:2008jh,Bechtle:2011sb,Bechtle:2013wla} and  \HS\ (version 1.4.0)~\cite{Bechtle:2013xfa}. 
A summary of all constraints on the maximal mixing angle $\sin\al$ is shown in Fig. ~\ref{fig:sinamw}.
\begin{figure}
 \includegraphics[width=0.8\textwidth]{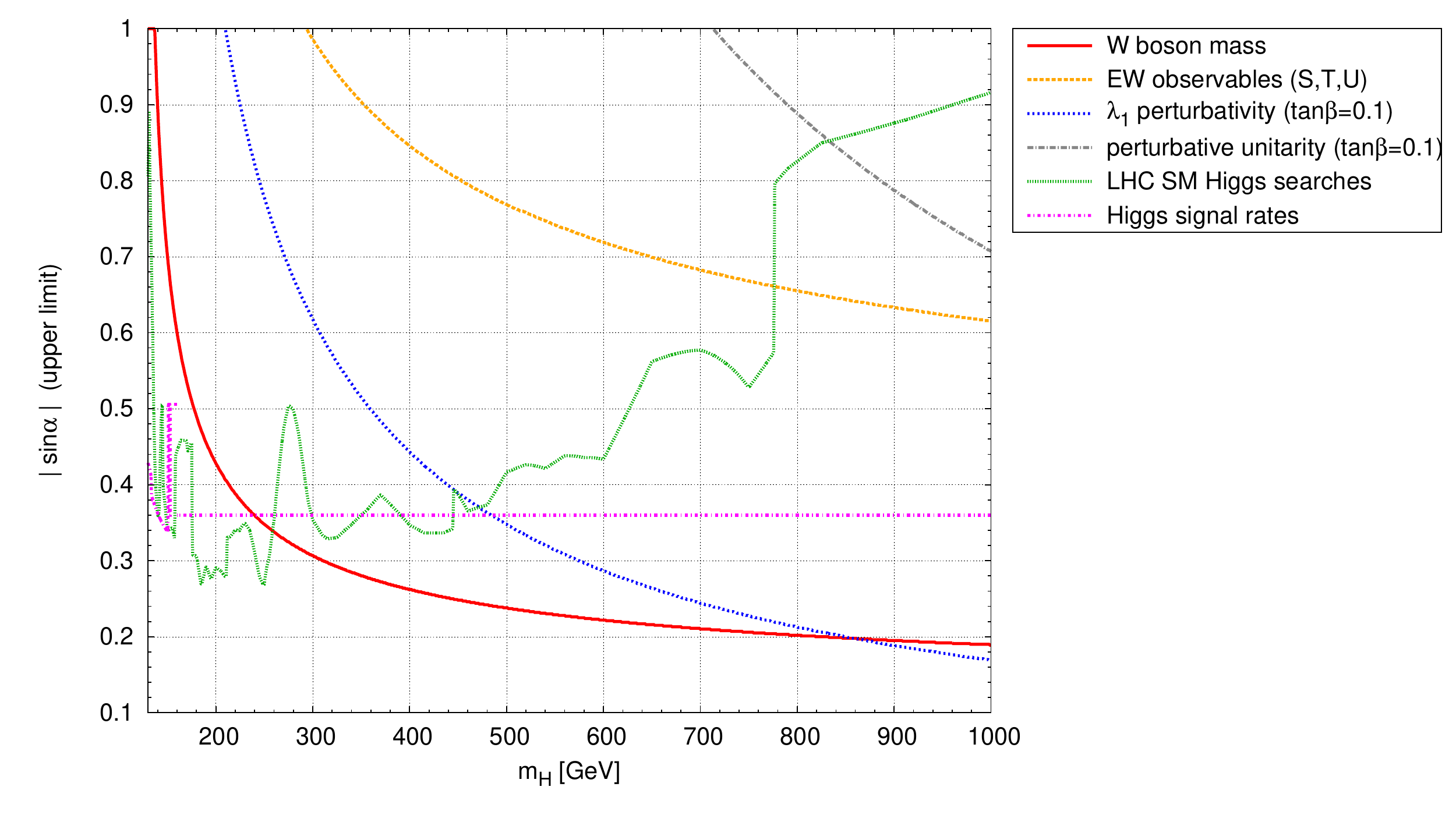}
\caption{\label{fig:sinamw} Maximal allowed values for $| \sin\al |$ in the high mass region, $m_H\in [130, 1000]\,\GeV$, from  {NLO} calculations of the $W$ boson mass (\emph{red, solid})~\cite{Lopez-Val:2014jva}, electroweak precision observables (EWPOs) tested via the oblique parameters $S$, $T$ and $U$ (\emph{orange, dashed}), perturbativity of the RG-evolved coupling $\lam_1$ (\emph{blue, dotted}), evaluated {for an exemplary choice} $\tan\be\,=\,0.1$, perturbative unitarity (\emph{grey, dash-dotted}), direct LHC Higgs searches (\emph{green, dashed}), and the Higgs signal strength (\emph{magenta, dash-dotted}). Taken from \cite{Robens:2016xkb}.}
\end{figure}
{Production} cross-sections for the 14 \TeV~ LHC, after all constraints have been taken into account, are shown in Fig. \ref{fig:xsecs} for the high-mass range. Specific benchmarks for all mass ranges have been presented in \cite{Robens:2016xkb}\footnote{See also \cite{YR4}.}.

 \begin{figure}
 \centering
 \subfigure[~{Heavy Higgs signal rate with SM particles in the final state for the LHC at $14~\TeV$}.]{
 \includegraphics[width=0.38\textwidth]{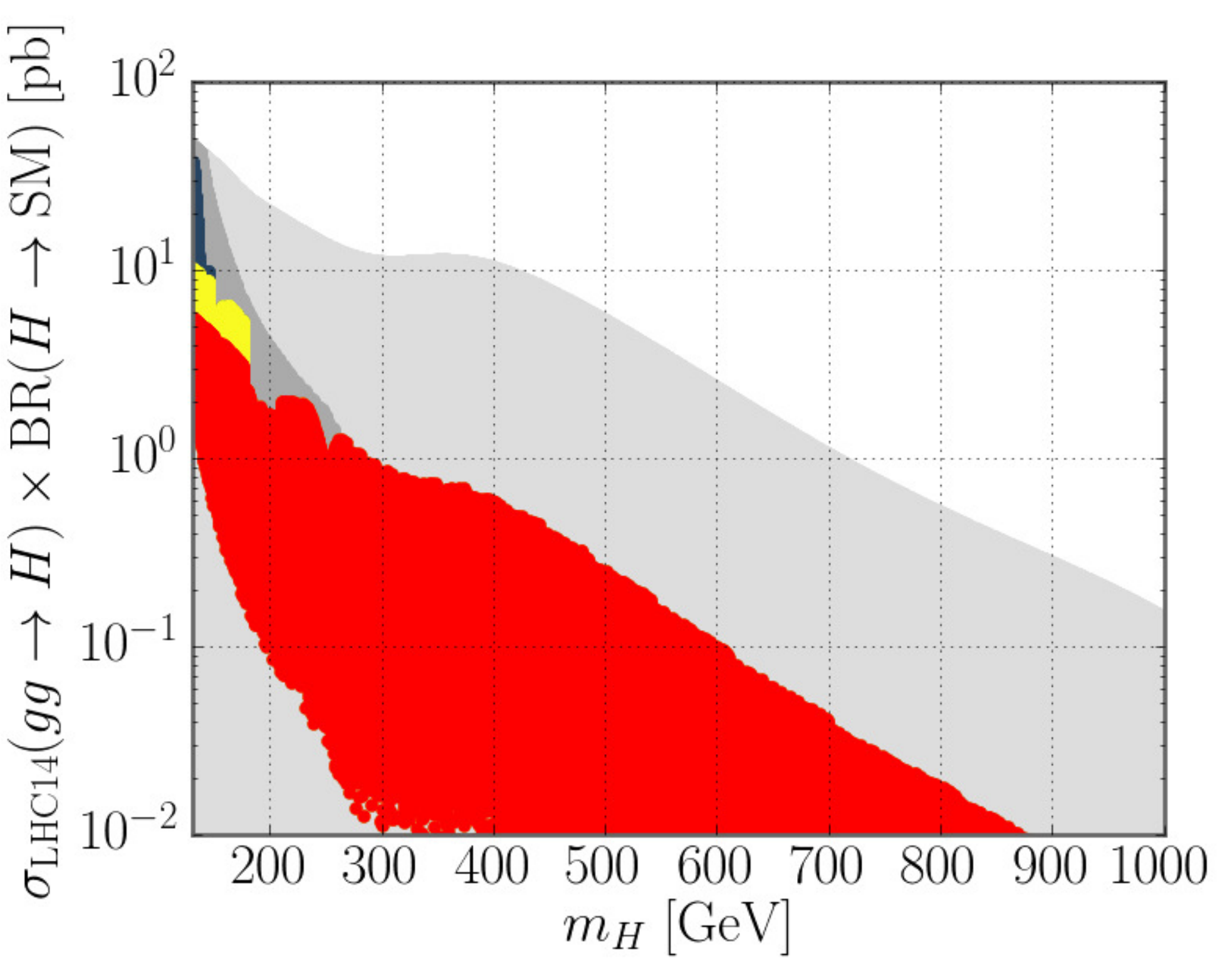}
 }
 \hfill
 \subfigure[~{Heavy Higgs signal rate with light Higgs bosons in the final state for the LHC at $14~\TeV$}.]{
 \includegraphics[width=0.38\textwidth]{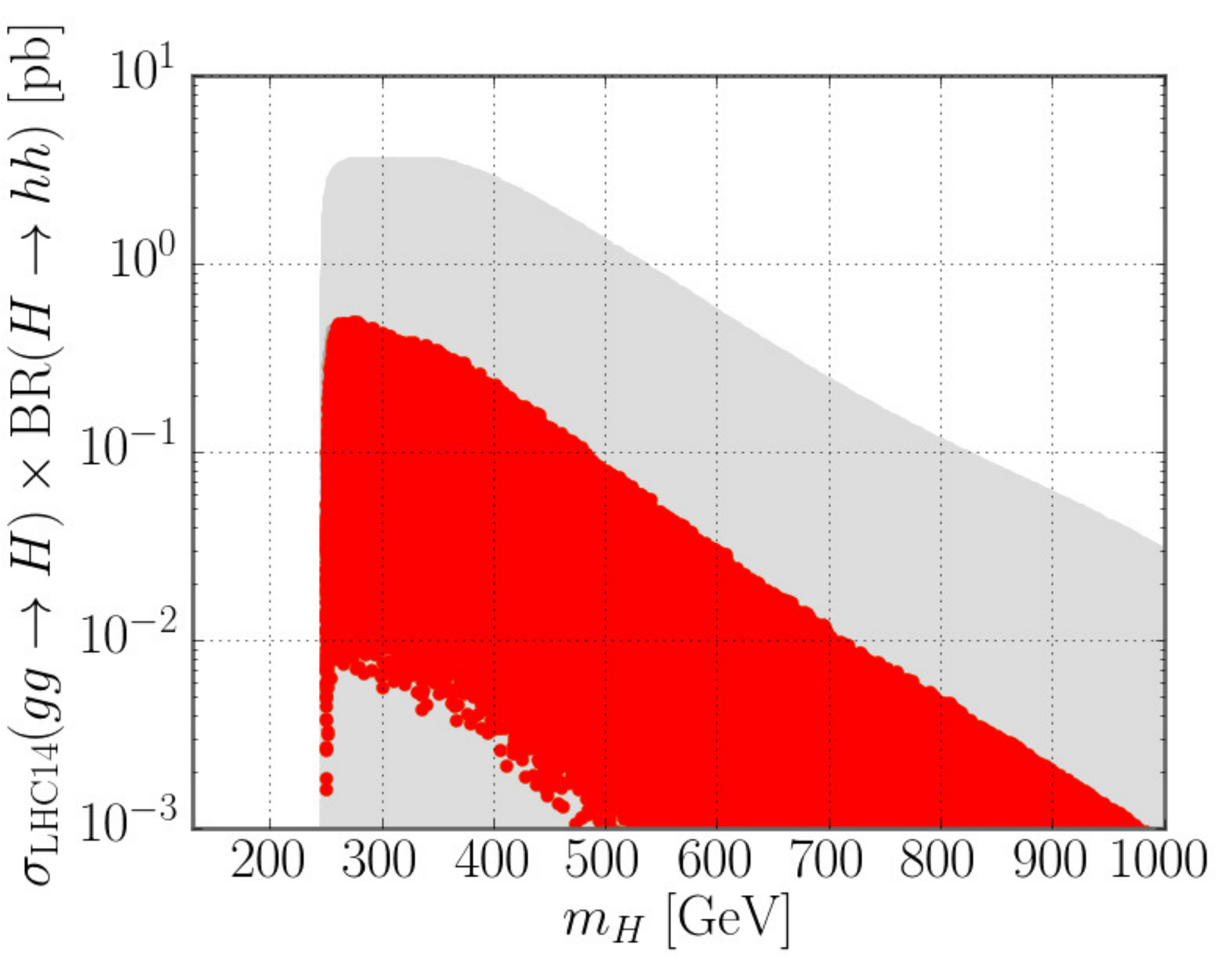}
 }
 \caption{\label{fig:xsecs} Production cross-sections at a 14 \TeV~ LHC, for a heavy Higgs $H$ decaying into SM particles {\sl (left)} or $hh$ final states {\sl (right)}; for the latter, electroweak corrections have not been included. Cross sections stem from a simple rescaling of production cross sections presented in \cite{YR4}. Red and yellow regions correspond to agreement with the Higgs signal strength measurements at the $1\sigma$ and  $2\sigma$ level, respectively, blue points comply with direct experimental searches but do not agree with the Higgs signal strength within $2\sigma$. Taken from \cite{Robens:2016xkb}.}
 \end{figure}

\section{Renormalization}
The complete electroweak renormalization of the singlet model has been presented in \cite{Bojarski:2015kra}, and we refer the reader to this reference for explicit details. Here we only want to point to two major features of our scheme setup.\\

\noindent
{\bf Non-linear gauge fixing} We use a non-linear gauge fixing, specified by
\begin{equation}
\label{gaugefixing}
{\mathscr L}_{GF} = -\frac{1}{\xi_W} F^+ F^- - \frac{1}{2  {{ \xi_Z}}  }|
F^Z|^2 - \frac{1}{2 {{ \xi_A}} } | F^A|^2\, ,
\end{equation}\noi 
where the functions $F$ depend non-linearly on the Higgs and gauge fields and are given by Eqns. (21)-(23) of \cite{Bojarski:2015kra}. The gauge-fixing terms explicitly depend on
the non-linear gauge-fixing quantities $\tilde{\delta}_i$. We perform our implementation of the singlet model 
using 
{\sc Sloops} (see e.g. 
\cite{Boudjema:2005hb,Baro:2009gn}). \\

\noindent
{\bf Gauge-parameter independent physical results} We have {studied different schemes and} explicitly tested gauge-fixing parameter dependence. 
An improved On-shell prescription leads to gauge-parameter independent predictions for the one-loop corrections {to $\Gamma_{H\,\rightarrow\,h\,h}$:} 
\begin{equation}
\label{eq:improvedOS}
 \delta m^2_{hH} = 
\mbox{Re}\,\Sigma_{hH}(\pstar^2)\big|_{\xi_{W}=\xi_Z=1,\tilde\delta_i=0} 
\quad \mbox{with}\quad\pstar^2 = \frac{m_h^2+m_H^2}{2}\, ,
\end{equation}\noindent
This prescription coincides with the discussion in \cite{Baro:2009gn} in the context of supersymmetry{, and can also be related to the so-called pinch technique (see e.g. }\cite{Espinosa:2002cd}). 

We rely on two independent implementations of the model\footnote{See \cite{Bojarski:2015kra} for a complete description of the computational setup.}. Once all present constraints on the model are included, we find mild
NLO corrections, typically of few percent, and with theoretical uncertainties on the per mille level. Sample {results} for the one-loop eletroweak corrections to the decay width $\Gamma_{H\,\rightarrow\,h\,h}$ are displayed in Fig. \ref{fig:nlo}.

\begin{figure}
 \centering
 \subfigure[~$m_h\,=\,125.09\GeV$]{
 \includegraphics[width=0.38\textwidth]{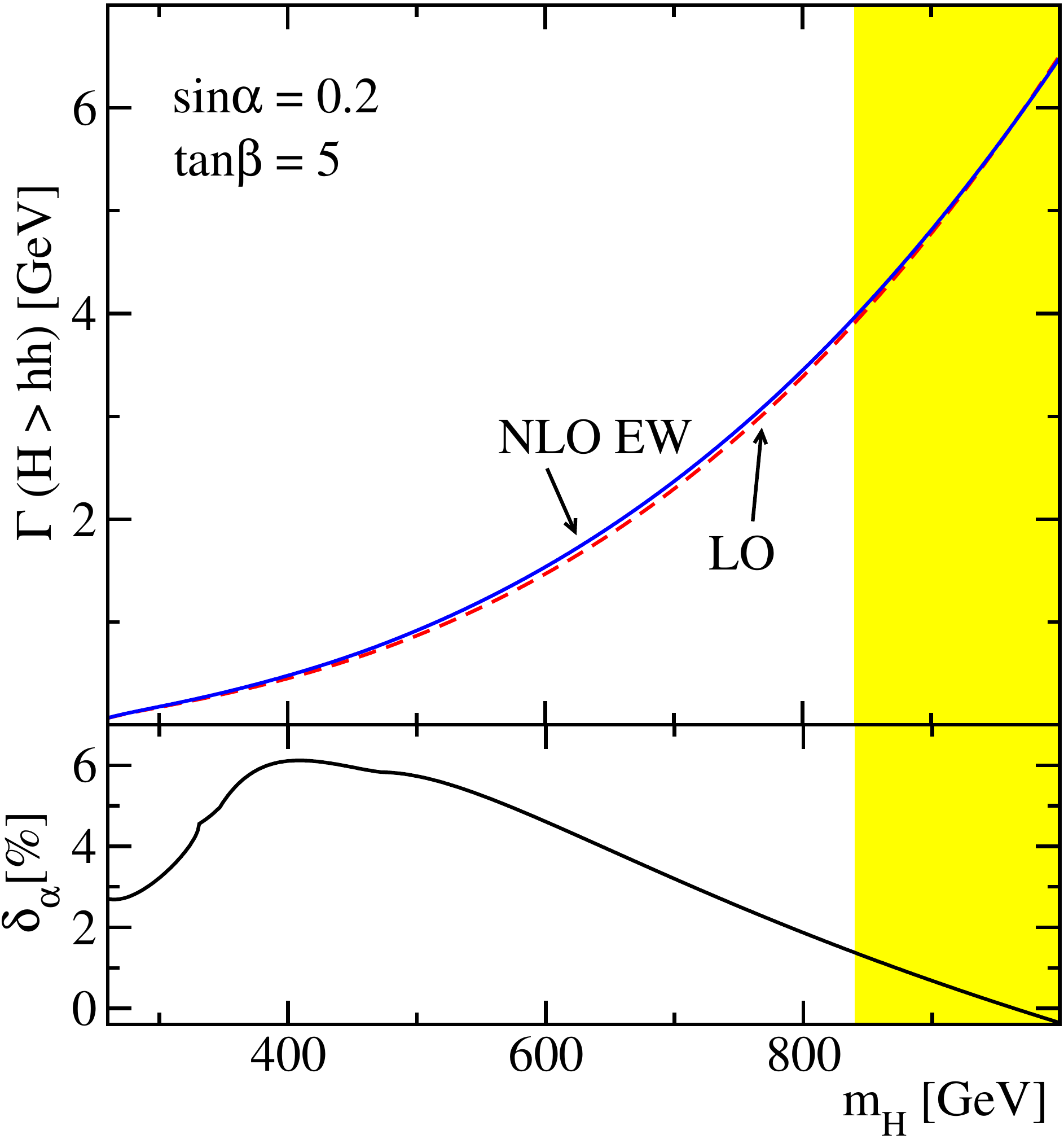}}
 \hfill
 \subfigure[~$m_H\,=\,125.09\GeV$]{
 \includegraphics[width=0.42\textwidth]{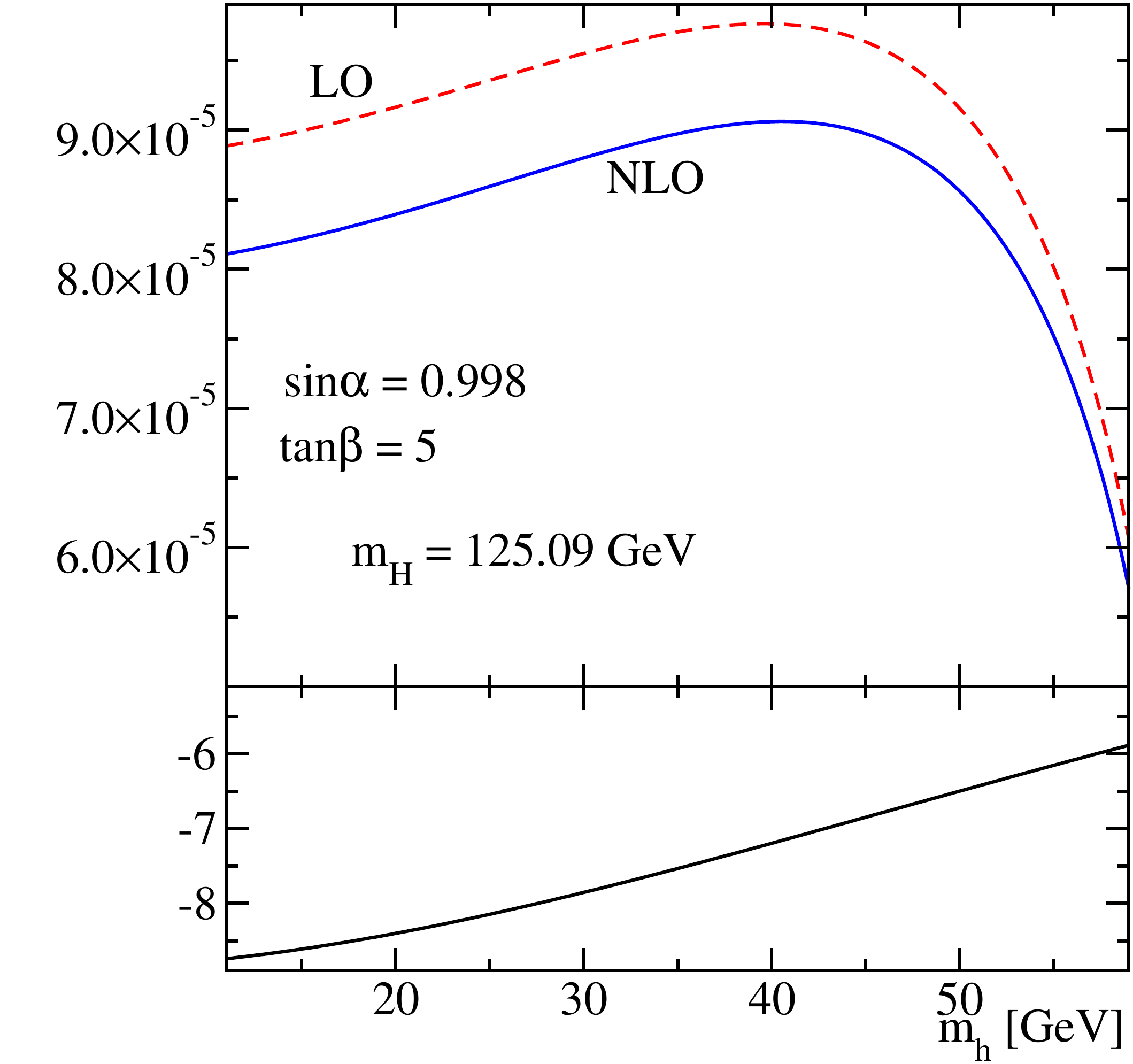}}
\caption{\label{fig:nlo} NLO corrections to the $H\,\rightarrow\,h\,h$ partial decay width, for fixed $\sin\al,\,\tan\be$ values and $m_h$ {\sl (left)} or $m_H$ {\sl (right)} being the 125 \GeV~resonance measured
at the LHC, as a function of the second scalar mass. We display the total decay width for $H\,\rightarrow\,h\,h$, as well as the {\sl relative} correction in the $\al_\text{em}$ input scheme for the electroweak parameters (see \cite{Bojarski:2015kra} for details). The yellow region is excluded {by} perturbativity of the couplings. {\sl Note:} $\tan\be$ is defined as $\frac{v_s}{v}$ in this case, in contrast to the definitions given above. Taken from \cite{Bojarski:2015kra}.}
\end{figure}

\section*{Acknowledgements}
\vspace{-3mm}
TR and TS want to thank M. Gouzevitch and  M. Slawinska for a useful discussion regarding one of the figures presented in \cite{Robens:2016xkb}.
\footnote{ Funding is acknowledged from the F.R.S.-FNRS "Fonds de la Recherche Scientifique", the Theory- LHC-France initiative
of CNRS/IN2P3, the U.S.~Department of Energy grant number DE-SC0010107 and the Alexander von Humboldt foundation.}

\begin{thebibliography}{10}

\bibitem{Schabinger:2005ei}
R.~Schabinger and J.~D. Wells, \emph{{A Minimal spontaneously broken hidden
  sector and its impact on Higgs boson physics at the large hadron collider}},
  \href{http://dx.doi.org/10.1103/PhysRevD.72.093007}{\emph{Phys.Rev.} {\bf
  D72} (2005) 093007}, [\href{http://arxiv.org/abs/hep-ph/0509209}{{\tt
  hep-ph/0509209}}].

\bibitem{Patt:2006fw}
B.~Patt and F.~Wilczek, \emph{{Higgs-field portal into hidden sectors}},
  \href{http://arxiv.org/abs/hep-ph/0605188}{{\tt hep-ph/0605188}}.

\bibitem{Bowen:2007ia}
M.~Bowen, Y.~Cui and J.~D. Wells, \emph{{Narrow trans-TeV Higgs bosons and H
  $\rightarrow$ hh decays: Two LHC search paths for a hidden sector Higgs
  boson}}, \href{http://dx.doi.org/10.1088/1126-6708/2007/03/036}{\emph{JHEP}
  {\bf 0703} (2007) 036}, [\href{http://arxiv.org/abs/hep-ph/0701035}{{\tt
  hep-ph/0701035}}].

\bibitem{Pruna:2013bma}
G.~M. Pruna and T.~Robens, \emph{{The Higgs Singlet extension parameter space
  in the light of the LHC discovery}},
  \href{http://dx.doi.org/10.1103/PhysRevD.88.115012}{\emph{Phys.Rev.} {\bf
  D88} (2013) 115012}, [\href{http://arxiv.org/abs/1303.1150}{{\tt
  1303.1150}}].

\bibitem{Robens:2015gla}
T.~Robens and T.~Stefaniak, \emph{{Status of the Higgs Singlet Extension of the
  Standard Model after LHC Run 1}},
  \href{http://dx.doi.org/10.1140/epjc/s10052-015-3323-y}{\emph{Eur. Phys. J.}
  {\bf C75} (2015) 104}, [\href{http://arxiv.org/abs/1501.02234}{{\tt
  1501.02234}}].

\bibitem{Robens:2016xkb}
T.~Robens and T.~Stefaniak, \emph{{LHC Benchmark Scenarios for the Real Higgs
  Singlet Extension of the Standard Model}},
  \href{http://dx.doi.org/10.1140/epjc/s10052-016-4115-8}{\emph{Eur. Phys. J.}
  {\bf C76} (2016) 268}, [\href{http://arxiv.org/abs/1601.07880}{{\tt
  1601.07880}}].

\bibitem{Lopez-Val:2014jva}
D.~Lopez-Val and T.~Robens, \emph{{Delta r and the W-boson mass in the Singlet
  Extension of the Standard Model}},
  \href{http://dx.doi.org/10.1103/PhysRevD.90.114018}{\emph{Phys.Rev.} {\bf
  D90} (2014) 114018}, [\href{http://arxiv.org/abs/1406.1043}{{\tt
  1406.1043}}].

\bibitem{ATLAS-CONF-2015-044}
{ATLAS and CMS Collaborations}, \emph{{Measurements of the Higgs boson
  production and decay rates and constraints on its couplings from a combined
  ATLAS and CMS analysis of the LHC pp collision data at $\sqrt{s}$ = 7 and 8
  TeV}}, .

\bibitem{Bechtle:2008jh}
P.~Bechtle, O.~Brein, S.~Heinemeyer, G.~Weiglein and K.~E. Williams,
  \emph{{HiggsBounds: Confronting Arbitrary Higgs Sectors with Exclusion Bounds
  from LEP and the Tevatron}},
  \href{http://dx.doi.org/10.1016/j.cpc.2009.09.003}{\emph{Comput.~Phys.~Commu%
n.} {\bf 181} (2010) 138}, [\href{http://arxiv.org/abs/0811.4169}{{\tt
  0811.4169}}].

\bibitem{Bechtle:2011sb}
P.~Bechtle, O.~Brein, S.~Heinemeyer, G.~Weiglein and K.~E. Williams,
  \emph{{HiggsBounds 2.0.0: Confronting Neutral and Charged Higgs Sector
  Predictions with Exclusion Bounds from LEP and the Tevatron}},
  \href{http://dx.doi.org/10.1016/j.cpc.2011.07.015}{\emph{Comput.~Phys.~Commu%
n.} {\bf 182} (2011) 2605}, [\href{http://arxiv.org/abs/1102.1898}{{\tt
  1102.1898}}].

\bibitem{Bechtle:2013wla}
P.~Bechtle, O.~Brein, S.~Heinemeyer, O.~St{\aa}l, T.~Stefaniak et~al.,
  \emph{{HiggsBounds-4: Improved Tests of Extended Higgs Sectors against
  Exclusion Bounds from LEP, the Tevatron and the LHC}},
  \href{http://dx.doi.org/10.1140/epjc/s10052-013-2693-2}{\emph{Eur.~Phys.~J.~%
C} {\bf 74} (2013) 2693}, [\href{http://arxiv.org/abs/1311.0055}{{\tt
  1311.0055}}].

\bibitem{Bechtle:2013xfa}
P.~Bechtle, S.~Heinemeyer, O.~St{\aa}l, T.~Stefaniak and G.~Weiglein,
  \emph{{$HiggsSignals$: Confronting arbitrary Higgs sectors with measurements
  at the Tevatron and the LHC}},
  \href{http://dx.doi.org/10.1140/epjc/s10052-013-2711-4}{\emph{Eur.Phys.J.}
  {\bf C74} (2014) 2711}, [\href{http://arxiv.org/abs/1305.1933}{{\tt
  1305.1933}}].

\bibitem{YR4}
{The LHC Higgs Cross Section Working Group}, \emph{{CERN Yellow Report 4}}, .

\bibitem{Bojarski:2015kra}
F.~Bojarski, G.~Chalons, D.~Lopez-Val and T.~Robens, \emph{{Heavy to light
  Higgs boson decays at NLO in the Singlet Extension of the Standard Model}},
  \href{http://dx.doi.org/10.1007/JHEP02(2016)147}{\emph{JHEP} {\bf 02} (2016)
  147}, [\href{http://arxiv.org/abs/1511.08120}{{\tt 1511.08120}}].

\bibitem{Boudjema:2005hb}
F.~Boudjema, A.~Semenov and D.~Temes, \emph{{Self-annihilation of the
  neutralino dark matter into two photons or a Z and a photon in the MSSM}},
  \href{http://dx.doi.org/10.1103/PhysRevD.72.055024}{\emph{Phys. Rev.} {\bf
  D72} (2005) 055024}, [\href{http://arxiv.org/abs/hep-ph/0507127}{{\tt
  hep-ph/0507127}}].

\bibitem{Baro:2009gn}
N.~Baro and F.~Boudjema, \emph{{Automatised full one-loop renormalisation of
  the MSSM II: The chargino-neutralino sector, the sfermion sector and some
  applications}},
  \href{http://dx.doi.org/10.1103/PhysRevD.80.076010}{\emph{Phys. Rev.} {\bf
  D80} (2009) 076010}, [\href{http://arxiv.org/abs/0906.1665}{{\tt
  0906.1665}}].

\bibitem{Espinosa:2002cd}
J.~R. Espinosa and Y.~Yamada, \emph{{Scale independent and gauge independent
  mixing angles for scalar particles}},
  \href{http://dx.doi.org/10.1103/PhysRevD.67.036003}{\emph{Phys. Rev.} {\bf
  D67} (2003) 036003}, [\href{http://arxiv.org/abs/hep-ph/0207351}{{\tt
  hep-ph/0207351}}].

\end{thebibliography}
\providecommand{\href}[2]{#2}\begingroup\raggedright\endgroup

\end{document}